\documentclass[nofootinbib,showpacs,preprintnumbers,amsmath,amssymb,floatfix]
{revtex4}
\usepackage{graphicx}

\begin{document}


\title{Structure Function $F_1$ singlet in Double-Logarithmic Approximation}

\vspace*{0.3 cm}

\author{B.I.~Ermolaev}
\affiliation{Ioffe Physico-Technical Institute, 194021
 St.Petersburg, Russia}
\author{S.I.~Troyan}
\affiliation{St.Petersburg Institute of Nuclear Physics, 188300
Gatchina, Russia}

\begin{abstract}
The conventional ways to calculate the perturbative component of  the DIS structure function $F_1$ singlet
involve
approaches based on BFKL which account for the single-logarithmic contributions accompanying the
Born factor $1/x$. In contrast,
we account for the double-logarithmic (DL) contributions unrelated to $1/x$ and because of that they
were disregarded as negligibly small. We
calculate $F_1$ singlet in the Double-Logarithmic
Approximation (DLA) and account at the same time for the running $\alpha_s$ effects.
We start with total resummation of both quark and gluon DL contributions and obtain  the explicit expression
for $F_1$ in DLA. Then, applying the saddle-point method, we calculate the small-$x$ asymptotics
of $F_1$, which proves
to be of the Regge form with the leading singularity $\omega_0 = 1.066$.  Its  large value compensates for the
lack of the factor $1/x$ in the DLA contributions. Therefore, this Reggeon can be named
a new Pomeron which can be
quite important for description of all QCD processes involving the vacuum (Pomeron) exchanges at very
high energies.
We prove that the expression for the small-$x$ asymptotics of $F_1$ scales: it depends on a single variable $Q^2/x^2$
only instead of $x$ and $Q^2$ separately.
Finally,  we
show that the small-$x$ asymptotics reliably represent $F_1$ at $x \leq 10^{-6}$.
\end{abstract}


\maketitle

\section{Introduction}

Description of the structure function $F_1$ singlet in the framework of Collinear Factorization usually involves DGLAP\cite{dglap}
to calculate the perturbative contributions. In this case $F_1$  is represented in the form of two convolutions:

\begin{equation}\label{f1dglap}
F_1 = C_q (x/y) \otimes \Delta q (y,Q^2)+ C_g (x/y) \otimes
\Delta g (y,Q^2),
\end{equation}
where $C_q$ and $C_g$ are the coefficient functions  and $\Delta q $ and $\Delta g$ denote the
evolved (with respect to $Q^2$) quark and gluon distributions respectively.  These distributions are solutions to the DGLAP equations
which govern the
$Q^2$-evolution of the initial quark and gluon distributions $\delta q (x,\mu^2)$ and $\delta g(x,\mu^2)$, evolving them
from the scale $\mu^2$ to $Q^2$. Both  $\delta q $ and $\delta g $ are defined at
$x \sim 1$ and $Q^2 = \mu^2 \sim 1$GeV$^2$.  The parameter $\mu$  is also called the factorization scale.
The $x$-dependence of $F_1$ is described by  the coefficient functions $C_{q,g}$ as well as by the phenomenological factors in
$\delta q, \delta g$. In the framework of DGLAP the evolution in the $k_{\perp}$-space is
is separated from evolution with respect to $x$. Such a separation takes place at $x \sim 1$ only and
breaks at small $x$ as was shown in Ref.~\cite{ggfl}. It is the theoretical reason not to use DGLAP at small $x$.
A practical reason is that DGLAP, by its design, accounts for the total
resummation of $\ln^n Q^2$ while contributions $\sim \ln^n x$ are present in
the DGLAP expressions in few first
orders in $\alpha_s$ only (through the coefficient functions in NLO,NNLO, etc.).

On the other hand,
such contributions are very important at small $x$, so it would be appropriate to
substitute the DGLAP expressions for the DIS structure functions by new ones which include the total resummation of
all double-logarithmic (DL) contributions. In the first place there are DL terms $\sim (\alpha_s \ln^2 (1/x))^n$, then
the terms $\sim (\alpha_s \ln (1/x)\ln Q^2)^n$, etc. Expressions accounting for resummation of DL contributions and
for the running $\alpha_s$ effects were
obtained for several
 structure functions with non-vacuum exchanges in the $t$-channel:
 the spin structure function $g_1$ (the singlet and non-singlet components) and the non-singlet component of $F_1$ (see
the overview\cite{egtg1sum} and refs therein).  Besides, there were obtained the expressions for $g_1$ and
non-singlet $F_1$ combining the DGLAP
results and resummation of the DL contributions, which made possible to apply these expressions at arbitrary $x$ and $Q^2$.

However, a similar generalization of DGLAP
was not obtained for the singlet $F_1$.
The point is that
by that time $F_1$ in the small-$x$ region has been intensively investigated in terms of approaches based on BFKL\cite{bfkl} and this looked as the only way
to study $F_1$ at small $x$. Indeed, the leading $x$-dependent contributions to
$F_1$  proved to be  the single-logarithmic (SL) terms accompanying the "Born" factor $1/x$:

\begin{equation}\label{bfklser}
(1/x) \left[1 + c_1 \alpha_s \ln (1/x) + c_2 (\alpha_s \ln (1/x))^2 + ...\right]
\end{equation}
while the DL contributions proportional to $1/x$ , i.e. the terms

\begin{equation}\label{dlcancel}
 (1/x) \left[1 + c^{DL}_1 \alpha_s \ln^2 (1/x) + c_2^{DL} (\alpha_s \ln (1/x))^4 + ...\right],
\end{equation}
cancel each other (i.e. $c_k^{DL} = 0$ for $k= 1,2,..$) as was found first in Ref.~\cite{ttwu}.
As a result, the common strategy for investigating the QCD processes with vacuum exchanges in the $t$-channel
was based on the use of the BFKL results. In particular, SL contributions to the structure functions
$F_{1,2}$ was presented in Refs.~\cite{catciaf,cathaut};
SL contributions to $F_2$ in Ref.~\cite{ehb,haut} were calculated with inclusion of resummed anomalous dimensions in the renormalization group
equation
while $F_2$  in Ref.~\cite{kms} was calculated with direct unification
of DGLAP and FFKL.

Solution to the BFKL equation is expressed through the series of the high-energy
asymptotics of the Regge form, with the leading
asymptotics commonly addressed as the BFKL Pomeron, so at $x \to 0$

\begin{equation}\label{f1bfkl}
F_1 \sim x^{-(1 + \Delta_P)},
\end{equation}
where  $\Delta_P$ is the Pomeron intercept.  As $\Delta_P > 0$ for the both LO and NLO BFKL Pomerons, they are
called the supercritical ones. As we are not going to use BFKL or its modifications like \cite{balkov} in the present paper, we just mention that
the extensive literature on this issue can be found in Ref.~\cite{iancu}.

Instead of using the BFKL results or trying to increase the accuracy of the method of Ref.~\cite{cathaut}, in the present paper we
account for total resummation of the double-logarithmic contributions to $F_1$. In the first place we account for the
$x$-dependent contributions

\begin{equation}\label{dlser}
1 + c'_1 \alpha_s \ln^2 (1/x) + c'_2 (\alpha_s \ln^2 (1/x))^2 + ...
\end{equation}
and then for DL terms combining logs of $x$ and $Q^2$. These DL contributions
do not involve the large factor $1/x$ and by this reason they have been neglected
in the BFKL approach.
We calculate the singlet structure function $F_1$ in DLA, summing DL contributions coming from virtual gluon and quark
exchanges. As a result, our expressions for coefficient functions and anomalous dimensions contain total resummations of
appropriate DL terms.
To calculate $F_1$ we compose and solve Infra-Red Evolution Equations (IREE)
in the same way as we did for calculating the DIS structure function $g_1$ singlet (see Ref.~\cite{egtg1sum}),
investigating the cases of fixed and running $\alpha_s$.
We remind that the IREE method was suggested by L.N.~Lipatov in Ref.~\cite{kl}. It is based on factorization of DL contributions of the partons with minimal
transverse momenta first noticed by V.N.~Gribov in Ref.~\cite{grib} in the context of QED of hadrons.
Technology of implementation of this method to DIS is described in detail in Ref.~\cite{egtg1sum}.
In contrast to DGLAP and BFKL equations, we compose the two-dimensional evolution equations: They control evolutions in both $x$ and $Q^2$.
We obtain the explicit expression for $F_1$ and then, applying the saddle-point method, we calculate
the small-$x$ asymptotics of $F_1$ automatically complemented by the
asymptotic $Q^2$-dependence. The asymptotics proves to be of the Regge form. The large value of  the intercept
compensates for the lack of the factor $1/x$ in the DL contributions and thereby makes the DLA asymptotics
be of the same order as
the BFKL one. This proves that the DL contributions to $F_1$ at small $x$ are, at least, no less important
than the contributions coming from the BFKL Pomeron.

Our paper is outlined as follows: in Sect.~II we compose and solve IREE for the Compton amplitudes $A_{q,g}$ related to  $F_1$
by the Optical theorem. In this Sect. we express $A_{q,g}$ through the amplitudes of the $2 \to 2$ scattering of partons.
Those amplitudes are calculated in Sect.~III. In Sect.~IV we apply the saddle-point method to obtain
explicit expression for
the small-$x$ asymptotics of $F_1$ and prove that this asymptotics depends on the single variable $Q^2/x^2$ instead of
separate dependence on $Q^2$ and $x$.
In Sect.~V  we consider in detail the intercept of the Pomeron in DLA, embracing
the cases of fixed and running $\alpha_s$. We also fix the region where the small-$x$ asymptotics can reliably
represent $F_1$. Finally, Sect.~VI is for our concluding remarks.

\section{IREE for the amplitudes of Compton scattering off partons}

Following the DGLAP pattern, we consider $F_1$ in the framework of Collinear Factorization and represent $F_1$
through the convolutions of the perturbative components $T_q$ and $T_g$ with non-perturbative initial quark and gluon distributions
$\phi_{q,g}$ respectively:

\begin{equation}\label{f1colfact}
F_1 = F^{q}_1 \otimes \phi_q + F_1^{g}g \otimes \phi_g.
\end{equation}

Throughout the paper we will consider the perturbative objects $F_1^{q,g}$ only.  It is convenient to consider the Compton
amplitudes $A_q$ and $A_g$ related to $T_1^{q,g}$
by Optical theorem:

\begin{equation}\label{opt}
F_1^{q,g} (x,Q^2/\mu^2) = -\frac{1}{2\pi} \Im A_{q,g}(x,Q^2/\mu^2),
\end{equation}
where we have introduced the factorization scale $\mu$ and used the standard notation $x = Q^2/w$, with $w = 2pq$ and $Q^2 = - q^2$.
The next step is to represent $A_{q,g}$ in terms of the Mellin transform:

 \begin{equation}\label{mellin}
A_{q,g}(w/\mu^2, Q^2/\mu^2) = \int_{- \imath \infty}^{\imath \infty}
\frac{d \omega}{2 \pi \imath} \left(w/\mu^2\right)^{\omega}\xi^{(+)}(\omega) F_{q,g}(\omega, Q^2/\mu^2)
\approx \int_{- \imath \infty}^{\imath \infty}
\frac{d \omega}{2 \pi \imath} e^{\omega \rho} F_{q,g}(\omega, y),
\end{equation}
where we have introduced the signature factor $\xi^{(+)}(\omega) = \left(1 + e^{- \imath \omega}\right)/2 \approx 1$ and
the logarithmic variables $\rho, y$ (using the standard notation $w = 2pq$):
\begin{equation}\label{y12}
\rho = \ln (w/\mu^2),~~y = \ln (Q^2/\mu^2).
 \end{equation}
 In what follows we will address $F_q, F_g$ as Mellin amplitudes and will use the same form of the Mellin transform for other amplitudes as well.
 For instance, the Mellin transform for the color singlet amplitude $A_{gg}$ of the elastic gluon-gluon scattering in the forward kinematics is

 \begin{equation}\label{agg}
 A_{gg} = \int_{- \imath \infty}^{\imath \infty}
\frac{d \omega}{2 \pi \imath} \left(w/\mu^2\right)^{\omega}\xi^{(+)}(\omega) f_{gg} (\omega) \approx
\int_{- \imath \infty}^{\imath \infty}
\frac{d \omega}{2 \pi \imath} e^{\omega \rho} f_{gg} (\omega).
 \end{equation}
We have presumed in Eq.~(\ref{agg}) that virtualities of all external gluons are $\sim \mu^2$.
Let us notice that the only difference between the Mellin representation for the Compton amplitudes $A_{q,g}$ and
the similar amplitudes related to the singlet $g_1$ is in
the signature factors only: the signature factor for $g_1$ is $\xi^{(-)}(\omega) = \left(-1 + e^{- \imath \omega}\right)/2$.
Otherwise, technology of composing and solving IREE for $A_{gg}$ and $g_1$ singlet is the same.
Because of that we present IREE for $F_q, F_g$ (and for auxiliary amplitudes as well) with short comments only.
The full-length derivation of all involved IREE can be found in
Ref.~\cite{egtg1sum}.
Now all set to construct IREEs for $F_{q,g}$.  In the kinematics where

\begin{equation}\label{kinem}
w \gg Q^2 \gg \mu^2,
\end{equation}
the amplitudes $F_q, F_g$ obey the partial differential equations:

 \begin{eqnarray}\label{eqfy}
\left[\partial/\partial y + \omega \right] F_{q}(\omega,y) &=&
F_{q} (\omega, y) h_{qq} (\omega) + F_{g} (\omega, y) h_{gq} (\omega),
\\ \nonumber
\left[\partial/\partial y + \omega \right] F_{g}(\omega, y) &=&
F_{q} (\omega, y) h_{qg} (\omega) + F_{g} (\omega, y) h_{gg} (\omega),
\end{eqnarray}
where we have used the following convenient notations:
\begin{equation}\label{fh}
h_{rr'} = \frac{1}{8 \pi^2} f_{rr'},
\end{equation}
with $r,r' = q,g$ and $f_{rr'}$ being the parton-parton amplitudes. We will calculate $h_{rr'}$ in the next Sect.
Actually, the equations in (\ref{eqfy}) manifest strong resemblance with the DGLAP equations. Indeed, the first factor in
brackets in the l.h.s. of (\ref{eqfy}) exists in DGLAP too. The second term vanishes when the Mellin factor $(s/\mu^2)^{\omega}$
is replaced by the factor $x^{-\omega}$ which is used in the DGLAP equations. When the parton amplitudes $f_{rr'}$
are in the Born approximation, Eq.~(\ref{eqfy}) coincides with the DGLAP equations. A general solution to Eq.~(\ref{eqfy}) is

\begin{eqnarray}\label{fy}
F_{q}(\omega,y) &=& e^{- \omega y}\left[C_{(+)} e^{\Omega_{(+)} y} + C_{(-)} e^{\Omega_{(-)} y}\right],
\\ \nonumber
F_{g}(\omega, y) &=& e^{- \omega y}\left[ C_{(+)} \frac{h_{gg} - h_{qq} + \sqrt{R}}{2h_{qg}} e^{\Omega_{(+)} y} +
C_{(-)} \frac{h_{gg} - h_{qq} - \sqrt{R}}{2h_{qg}} e^{\Omega_{(-)} y} \right],
\end{eqnarray}
where $C_{(\pm)} (\omega)$ are arbitrary factors whereas

\begin{equation}\label{omegapm}
\Omega_{(\pm)} = \frac{1}{2} \left[ h_{gg} + h_{qq} \pm \sqrt{R}\right]
\end{equation}
and

\begin{equation}\label{r}
R = (h_{gg} + h_{qq})^2 - 4(h_{qq}h_{gg} - h_{qg}h_{gq}) = (h_{gg} - h_{qq})^2  + 4 h_{qg}h_{gq} .
\end{equation}

We specify the factors $C_{(\pm)} (\omega)$  by the matching with the Compton amplitudes $f_q, f_g$ calculated in
the kinematics $Q^2 \approx \mu^2$, i.e. at $y = 0$. The matching condition is

\begin{equation}\label{match}
F_q(\omega,y)|_{y = 0} = f_q(\omega),~~F_g(\omega,y)|_{y = 0} = f_g(\omega),
\end{equation}
which leads to the following expressions:

\begin{eqnarray}\label{cpmf}
C_{(+)} &=& \frac{h_{qg} f_g(\omega) - \left(h_{gg} - h_{qq} -\sqrt{R}\right) f_q(\omega)}
{2 \sqrt{R}},
\\ \nonumber
C_{(-)} &=& \frac{ - h_{qg} f_g(\omega) + \left(h_{gg} - h_{qq} +\sqrt{R}\right) f_q(\omega)}
{2 \sqrt{R}}.
\end{eqnarray}

 Now let us express $f_q, f_g$ through the parton-parton
amplitudes $h_{rr'}$. To this end, we construct IREE for them. As $f_q, f_g$ do not depend on  $Q^2$, the IREE for them are algebraic:

\begin{eqnarray}\label{eqfy0}
 \omega f_q(\omega) &=&
a_{\gamma q} +
f_q (\omega) h_{qq} (\omega) + f_g (\omega) h_{gq} (\omega),
\\ \nonumber
\omega f_g(\omega) &=&
f_q (\omega) h_{qg} (\omega) + f_g (\omega) h_{gg} (\omega),
\end{eqnarray}
where $a_{\gamma q} = e^2$, with $e^2$ being the total electric charge of the involved quacks,
so that $a_{\gamma q}/\omega$ is the Born value of amplitude $f_q(\omega)$.
There is no a similar term in the equation for $f_g(\omega)$. The only difference between the r.h.s. of (\ref{eqfy0})
and (\ref{eqfy}) is the factor $a_{\gamma q}$ in Eq.~(\ref{eqfy0}).
 The solution to Eq.~(\ref{eqfy0}) is

\begin{eqnarray}\label{fy0}
 f_q (\omega) &=& a_{\gamma q} \frac{ (\omega - h_{gg})}{G(\omega)},
 \\ \nonumber
f_g (\omega) &=&  a_{\gamma q}\frac{ h_{qg}}{G(\omega)},
\end{eqnarray}
with $G(\omega)$ being the determinant of the system (\ref{eqfy0}):

\begin{equation}\label{det}
G = (\omega - h_{qq})(\omega - h_{gg})- h_{gg}h_{qg}.
\end{equation}

Combining Eqs.~(\ref{cpmf}) and (\ref{fy0}), we express $C_{\pm}$ through the parton-parton amplitudes:

\begin{eqnarray}\label{cpm}
C_{(+)} &=& a_{\gamma q}\frac{h_{qg}h_{gq} - (\omega - h_{gg})\left(h_{gg} - h_{qq} - \sqrt{R}\right)}{2 G \sqrt{R}},
\\ \nonumber
C_{(-)} &=& a_{\gamma q}\frac{ -h_{qg}h_{gq} + (\omega - h_{gg})\left(h_{gg} - h_{qq} + \sqrt{R}\right)}{2 G \sqrt{R}}.
\end{eqnarray}

Combining Eqs.~(\ref{cpm},\ref{omegapm}) and (\ref{fy}), we can easily express $F_{q,g}$ in terms of the parton-parton
amplitudes $h_{rr'}$.

\section{parton-parton amplitudes}

In this Sect. we obtain explicit expressions for the parton amplitudes $h_{rr'}$.
The IREE for $h_{rr'}$ are quite similar to Eq.~(\ref{eqfy0}):

\begin{eqnarray}\label{eqh}
\omega h_{qq} &=& b_{qq} + h_{qq}h_{qq} + h_{qg}h_{gq},~~\omega h_{qg} = b_{qg} + h_{qq}h_{qg}+ h_{qg}h_{gg},
\\ \nonumber
\omega h_{gq} &=& b_{gq} + h_{gq}h_{qq}+ h_{gg}h_{gq},~~\omega h_{gg} = b_{gg} + h_{gq}h_{qg}+ h_{gg}h_{gg},
\end{eqnarray}
where the terms $b_{rr'}$ include the Born factors $a_{rr'}$ and contributions of non-ladder graphs $V_{rr'}$:
\begin{equation}\label{bik}
b_{rr'} = a_{rr'} + V_{rr'}.
\end{equation}

The Born factors are (see Ref.~\cite{egtg1sum} for detail):

\begin{equation}\label{app}
a_{qq} = \frac{A(\omega)C_F}{2\pi},~a_{qg} = \frac{A'(\omega)C_F}{\pi},~a_{gq} = -\frac{A'(\omega)n_f}{2 \pi}.
~a_{gg} = \frac{2N A(\omega)}{\pi},
\end{equation}
where $A$ and $A'$ stand for the running QCD couplings:

\begin{eqnarray}\label{a}
A = \frac{1}{b} \left[\frac{\eta}{\eta^2 + \pi^2} - \int_0^{\infty} \frac{d z e^{- \omega z}}{(z + \eta)^2 + \pi^2}\right],
A' = \frac{1}{b} \left[\frac{1}{\eta} - \int_0^{\infty} \frac{d z e^{- \omega z}}{(z + \eta)^2}\right],
\end{eqnarray}
with $\eta = \ln \left(\mu^2/\Lambda^2_{QCD}\right)$ and $b$ being the first coefficient of the Gell-Mann- Low function. When the running effects for the QCD coupling
are neglected,
$A(\omega)$ and $A'(\omega)$ are replaced by $\alpha_s$. The terms $V_{rr'}$ are represented in a similar albeit more involved way (see Ref.~\cite{egtg1sum} for detail):

\begin{equation}
\label{vik} V_{rr'} = \frac{m_{rr'}}{\pi^2} D(\omega)~,
\end{equation}
with
\begin{equation}
\label{mik} m_{qq} = \frac{C_F}{2 N}~,\quad m_{gg} = - 2N^2~,\quad
m_{gq} = n_f \frac{N}{2}~,\quad m_{qg} = - N C_F~,
\end{equation}
and
\begin{equation}
\label{d} D(\omega) = \frac{1}{2 b^2} \int_{0}^{\infty} d z
e^{- \omega z} \ln \big( (z + \eta)/\eta \big) \Big[
\frac{z + \eta}{(z + \eta)^2 + \pi^2} - \frac{1}{z +
\eta}\Big]~.
\end{equation}

Let us note that $D = 0$ when the running coupling effects are neglected. It corresponds the total compensation of DL
contributions of non-ladder Feynman graphs to scattering amplitudes with the positive signature as was first
noticed in Ref.~\cite{nest}. When $\alpha_s$ is running, such compensation is only partial.
Solution to Eq.~(\ref{eqh}) is
\begin{eqnarray}\label{h}
&& h_{qq} = \frac{1}{2} \Big[ \omega - Z - \frac{b_{gg} -
b_{qq}}{Z}\Big],\qquad h_{qg} = \frac{b_{qg}}{Z}~, \\ \nonumber &&
h_{gg} = \frac{1}{2} \Big[ \omega - Z + \frac{b_{gg} -
b_{qq}}{Z}\Big],\qquad h_{gq} =\frac{b_{gq}}{Z}~,
\end{eqnarray}
where
\begin{equation}
\label{z}
 Z = \frac{1}{\sqrt{2}}\sqrt{ Y + W
}~,
\end{equation}
with
\begin{equation}\label{y}
Y = \omega^2 - 2(b_{qq} + b_{gg})
\end{equation}
and
\begin{equation}\label{w}
  W = \sqrt{(\omega^2 - 2(b_{qq} + b_{gg}))^2 - 4 (b_{qq} - b_{gg})^2 -
16b_{gq} b_{qg} }
\end{equation}

The algebraic equations (\ref{eqh}) are non-linear, so they yield four expressions for $Z$. We selected in Eq.~(\ref{z})
the solution obeying
the matching with the Born amplitudes $h^{Born}_{rr'}$: at large $\omega$

\begin{equation}\label{matchh}
h_{rr'} \to h^{Born}_{rr'} =  a_{rr'}/\omega.
\end{equation}

Substituting the expressions of Eq.~(\ref{h}) in (\ref{fy0}), we obtain explicit expressions for
amplitudes $f_q, f_g$. Combining them with
Eqs.~(\ref{cpm},\ref{omegapm}) and (\ref{fy}), we obtain explicit expressions for $F_q$ and $F_g$.
Substituting them in Eq.~(\ref{mellin}), we arrive at the explicit expressions for the  Compton amplitudes $A_q$ and $A_g$.
Finally, applying the Optical theorem (\ref{opt}) to $A_q$ and $A_g$, we  arrive at the structure function $F_1$ singlet.

\section{Small-$x$ asymptotics of the structure function $F_1$}

The regular way to obtain the small-$x$ asymptotics of  $A_q$ and $A_g$ is to write explicit expressions
for $F_q$ and $F_g$ in
Eq.~(\ref{mellin}), then push $x \to 0$ and apply the saddle-point method. However before doing this, let us consider
in derail how to calculate
the asymptotics of the gluon-gluon scattering amplitude $A_{gg}$, presuming virtualities of all external gluons $\sim \mu^2$.

\subsection{Asymptotics of $F_1$}

The small-$x$ asymptotics of $A_q$ and $A_g$ can be obtained
with applying the saddle-point method
to Eq.~(\ref{mellin}).  As $\Omega_{(+)} > \Omega_{(-)}$, we neglect
the terms $C_{(-)}$ in (\ref{fy}) and represent Eq.~(\ref{mellin}) to the following form:

 \begin{eqnarray}\label{aqgpsi}
 A_q &\approx&
\int_{- \imath \infty}^{\imath \infty}
\frac{d \omega}{2 \pi \imath} e^{\omega \xi} \widetilde{F}_q (\omega) e^{\Omega_{(+)} y} = \int_{- \imath \infty}^{\imath \infty}
\frac{d \omega}{2 \pi \imath} e^{\Psi_q},
\\ \nonumber
A_g &\approx&
\int_{- \imath \infty}^{\imath \infty}
\frac{d \omega}{2 \pi \imath} e^{\omega \xi} \widetilde{F}_g (\omega)
e^{\Omega_{(+)} y} = \int_{- \imath \infty}^{\imath \infty}
\frac{d \omega}{2 \pi \imath} e^{\Psi_g},
 \end{eqnarray}
with $\xi = \ln(1/x)$ and

\begin{equation}\label{tildefqg}
\widetilde{F}_q = C_{(+)},~~ \widetilde{F}_g = C_{(+)} \frac{\left(h_{gg} - h_{qq} + \sqrt{R}\right)}{2 h_{qg}}
\end{equation}
and

\begin{equation}\label{psi}
\Psi_q = \omega \xi + \ln \widetilde{F}_q,~~ \Psi_q = \omega \xi + \ln \widetilde{F}_q.
\end{equation}

The stationary point at $x \to 0$ of $\Psi_q$ is given by the rightmost root $\omega_0$
of the following equation:

\begin{equation}\label{omega0gen}
d \Psi_q/d\omega = \xi + \frac{\widetilde{F}'_q(\omega_0)}{\widetilde{F}_q (\omega_0)} = 0.
\end{equation}

When $\xi \to \infty$, it must be equated by some negative singular contribution in the second term of Eq.~(\ref{omega0gen}).
Using the explicit formulae for $\widetilde{F}_{q,g}$, one can conclude that such contribution comes from the
factor $1/W$. So, the stationary point $\omega_0$ is the rightmost root of the equation

\begin{equation}\label{eqomega0}
 (\omega^2 - 2b_{qq} -2 b_{gg})^2 - 4(b_{qq}- b_{gg})^2
- 16 b_{qg}b_{gq} = 0.
\end{equation}

We consider in detail solutions to Eq.~(\ref{eqomega0}) at fixed and running $\alpha_s$ in the next Sect.
In vicinity of $\omega_0$ we can represent Eq.~(\ref{omega0gen}) as

\begin{equation}\label{eqw0}
\Psi'_q = \xi + \frac{\partial \widetilde{F}_q}{\widetilde{F}_q \partial W} \frac{d W}{d \omega} =
\xi + \frac{\partial \widetilde{F}_q}{\widetilde{F}_q \partial W} \frac{\lambda}{W} = \xi -
\varphi \frac{\lambda}{W} = 0,
\end{equation}
with

\begin{equation}\label{phi}
\varphi_q = - \partial \ln \widetilde{F}_q/ \partial W
\end{equation}
 and

\begin{equation}\label{lambda}
\lambda = 2 \omega \left(\omega^2 - 2 (b_{+}) + b_{(-)}\right),
\end{equation}
so in vicinity of the singularity $\omega_0$

\begin{equation}\label{w0}
W \approx W_0 =
\varphi \frac{\lambda}{\xi}
\end{equation}
Expanding $\Psi_q$ in the series, we obtain

\begin{equation}\label{psiser}
\Psi_q (\omega) \approx \Psi_q (\omega_0) + (1/2) {\Psi'}'_q (\omega_0) (\omega - \omega_0)^2.
\end{equation}

In order to calculate ${\Psi'}'_q$ we notice that the most singular contributions comes from differentiation of the
numerator in Eq.~(\ref{eqw0}), so

\begin{equation}\label{psi2}
{\Psi'}'_q \approx - \frac{\partial \widetilde{F}_q}{\widetilde{F}_q \partial W}
\frac{\lambda d W^{-1}}{d \omega} =
\varphi \frac{\lambda^2}{W^3} = \frac{\xi^3}{\lambda \varphi_q^2}
\end{equation}
and therefore the asymptotics of $A_q$ at $x \to 0$ is

\begin{equation}\label{asaq}
A_q(x, Q^2/\mu^2) \sim A_q^{as}(x, Q^2/\mu^2) =
C_{(+)}(\omega_0) \varphi_q (\omega_0)~\sqrt{\frac{\lambda }{2\pi \xi^3}}
\left(\frac{1}{x}\right)^{\omega_0} \left(\frac{Q^2}{\mu^2}\right)^{\Omega_{(+)}(\omega_0)}.
\end{equation}

Repeating the reasoning above for $A_g$ and applying to them the
Optical theorem, we conclude that the small-$x$ asymptotics of $F_1$ is

\begin{equation}\label{asf1}
F_1 \sim \Pi(\omega_0, \xi) \left(\frac{1}{x}\right)^{\omega_0} \left(\frac{Q^2}{\mu^2}\right)^{\Omega_{(+)}(\omega_0)}
,
\end{equation}
where the factor $\Pi(\omega_0, \xi)$  is

\begin{equation}\label{pi}
\Pi (\omega_0, \xi) = C_{(+)}\sqrt{\frac{\lambda}{2 \pi \xi^3}} \left[\varphi_q \delta q + \varphi_g
\left(\frac{h_{gg}(\omega_0)- h_{qq}(\omega_0) + \sqrt{R(\omega_0)}}{2 h_{qg}(\omega_0)}\right)
\delta g  \right],
\end{equation}
with $\delta q$ and $\delta q$ being the initial quark and gluon densities. They do not include
singular factors $\sim x^{-a}$, with positive $a$. The Regge form of the asymptotics is brought entirely by
the perturbative contributions.
Let us notice that $\Pi \sim \ln^{-3/2} (1/x)$. Eq.~(\ref{asf1}) exhibits that the total resummation of DL
contributions leads to the Regge behavior of $F_1$ at small $x$.

\subsection{Asymptotic scaling}

Substituting the explicit expressions for $h_{rr'}$ of Eq.~(\ref{h}) in  Eq.~(\ref{omegapm}) and
using Eq.~(\ref{eqomega0}),  we obtain that $\Omega_{(+)}(\omega_0) = \omega_0/2$.
This allows us to write the asymptotics of $F_1$ of Eq.~(\ref{asf1}) in the following way:

\begin{equation}\label{scaling}
F_1 \sim \Pi(\omega_0, \xi) \left(\frac{1}{x}\right)^{\omega_0} \left(\frac{Q^2}{\mu^2}\right)^{\omega_0/2}
= \Pi (\omega_0, \xi) \left(\frac{Q^2}{x^2 \mu^2}\right)^{\omega_0/2}
\end{equation}

Eq.~(\ref{scaling}) manifests that $F_1(x, Q^2)$
at asymptotically high
energies
depends on the single variable $Q^2/x^2$  only. We name such confluence of the $x$ and $Q^2$ dependence
the asymptotic scaling. The same form of the asymptotic scaling was
obtained earlier for the structure function $g_1$ and the non-singlet component of $F_!$ (see Ref.~\cite{egtg1sum} for detail).
We stress that the asymptotic scaling for $F_1$ can be checked with analysis of available experimental data.
Moreover, $F_2 = 2x F_1$ at very small $x$, which proves the asymptotic scaling for $F_2$. Finally,
let us notice that the leading singularity $\omega_0$ in Eq.~(\ref{scaling}) does not depend on $Q^2$.

\section{Anatomy of the leading singularity $\omega_0$}

In this Sect. we consider in detail the leading singularity $\omega_0$ which is the rightmost root of Eq.~(\ref{eqomega0}).
In order to make the asymptotics of $F_1$ be looking similarly to Eq.~(\ref{f1bfkl}), we denote

\begin{equation}\label{delta}
\omega_0 = 1 + \Delta
\end{equation}
so that $\omega_0$ could look similarly to the BFKL leading singularity, see Eq.~(\ref{f1bfkl}).
Now let us discuss different scenarios for calculating $\omega_0$. In what follows we will address $\Delta$ as the DL Pomeron intercept.
We remind that in the straightforward  Reggeology concept
$\Delta = 0$ and the Pomeron with  $\Delta > 0$ is called the supercritical Pomeron.

\subsection{Intercept under approximation of fixed QCD coupling}

In the first place let us estimate $\omega_0$ for the case of fixed $\alpha_s$.
In this case DL contributions of non-ladder graphs totally cancel each other, so that $D = 0$ and
$b_{rr'} = a_{rr'}$, with $a_{rr'}$ defined in Eq.~(\ref{app}), where $A(\omega)$ and $A'(\omega)$
should be replaced by $\alpha_s$. Then the solution to Eq.~(\ref{eqomega0}) is

\begin{equation}\label{intafix}
\omega_0^{fix} =
\left(\frac{\alpha_s^{fix}}{\pi}\right)^{1/2} \left[4 N + C_F + \sqrt{(4N - C_F)^2 - 8 n_f C_F}\right]^{1/2}
\approx 2.63 \sqrt{\alpha_s},
\end{equation}
with the standard notations of the color factors $N = 3$, $C_F = (N^2-1)/(2N$) and $n_f=4$ is the flavour number.
According to Ref.~\cite{egtfix}, in this case $\alpha_s^{fix} \approx 0.24$ which gives $\omega_0^{fix} = 1.29$.
Using the representation of $\omega_0$ of Eq.~(\ref{delta}), we obtain

\begin{equation}\label{deltafix}
\Delta^{fix} =  \omega_0^{fix} - 1 = 0.29
\end{equation}
 which fairly
coincides with the well-known LO BFKL intercept $\Delta_{LO}$. However, $\Delta_{LO}$ corresponds to accounting
for gluon contributions only while $\Delta^{fix}$ accommodates both gluon and quark contributions. When
the quark contributions in Eq.~(\ref{intafix}) are dropped, the purely gluonic intercept $\Delta_g^{fix}$
becomes somewhat greater:
\begin{equation}\label{deltagfix}
\Delta_g^{fix} = 0.35
\end{equation}
which again bears a strong resemblance to the LO BFKL intercept. However, we are positive that the approximation of
fixed $\alpha_s$ can used for rough estimating only, so we
will not pursuit this approximation any longer.

\subsection{Intercept for the case of running coupling}

\begin{figure}[h]
  \includegraphics[width=.6\textwidth]{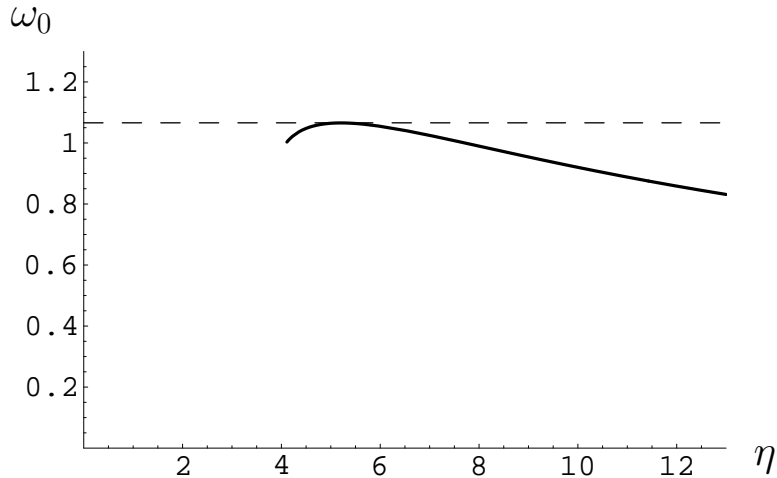}
  \caption{\label{f1dlfig1} Dependence of $\omega_0$ (upper solid curve) and $\omega_1$ (lower solid curve) on $\eta$~.
The dashing line shows maximum value of
   $\omega_0=1.066$ reached at $\eta=5.25$ that corresponds $\mu/\Lambda=13.8$.}
\end{figure}

Now we account for the running coupling effects in Eq.~(\ref{eqomega0}). Because of that, Eq.~(\ref{eqomega0})
can be solved only numerically.
As the couplings $A$ and $A'$ included in the factors $b_{rr'}$ depend on $\mu$ through
$\eta = \ln (\mu^2/\Lambda^2)$, the solution, $\omega_0$ is also $\mu$-dependent. Numerical calculations
yield the plot of the $\eta$-dependence of $\omega_0$ presented in Fig.~\ref{f1dlfig1}.
The curve in Fig.~\ref{f1dlfig1} has the maximum $\omega_0^{DLA} = 1.066$ at $\mu/\Lambda = 13.8$. We
address

\begin{equation}\label{muopt}
\mu_0 = 13.8 \Lambda
\end{equation}
as the optimal mass scale and call

\begin{equation}\label{deltadl}
\Delta = \omega_0^{DLA} - 1 = 0.066
\end{equation}
the intercept of the Pomeron in DLA.
It is interesting to notice that $\Delta$ is close to the NLO BFKL intercept. In contrast, when the quark contributions are neglected, the
purely gluonic intercept $\Delta_g^{DLA}$ is much greater:

\begin{equation}\label{deltadlg}
\Delta_g = 0.254~.
\end{equation}

Confronting Eq.~(\ref{deltagfix}) to (\ref{deltafix}) and  Eq.~(\ref{deltadlg}) to (\ref{deltadl}) demonstrates that accounting for
the quark contributions decreases the intercept. Similarly, confronting Eq.~(\ref{deltafix}) to (\ref{deltadl})
exhibits that accounting for the running
$\alpha_s$ effects essentially decreases the intercept value.
We also would like to stress that despite that our values of $\Delta$ in Eqs.~(\ref{deltagfix}) and (\ref{delta}) are close
to the values of the LO BFKL and NLO BFKL intercepts respectively, this similarity if just a coincidence: our intercepts are obtained from
resummation of DL contributions while the BFKL sums the single-logarithmic terms.  Moreover, Eq.~(\ref{deltadl}) corresponds to
the case of $\alpha_s$ running in every vertex of all involved Feynman graph while BFKL operates with fixed $\alpha_s$ and includes
setting of its scale a posteriori.

\subsection{Applicability region of the small-$x$ asymptotics}

It is obvious that the small-$x$ asymptotic expressions, like Eq.~(\ref{asf1}) are always much simpler than non-asymptotic expressions.
However, it is important to know at which values of $x$ the asymptotics can reliably be used. To answer this question we numerically investigate
 $R_{as}$ defined as follows:

\begin{equation}\label{ras}
R_{as} (x, Q^2) = \frac{A_q^{as}(x,Q^2)}{A_q (x,Q^2)}
\end{equation}

The $x$-dependence of $R_{as}$ at fixed $Q^2$ is shown in Fig.~2 for the case when $Q^2 \approx \mu^2$:

\begin{figure}[h]
  \includegraphics[width=.4\textwidth]{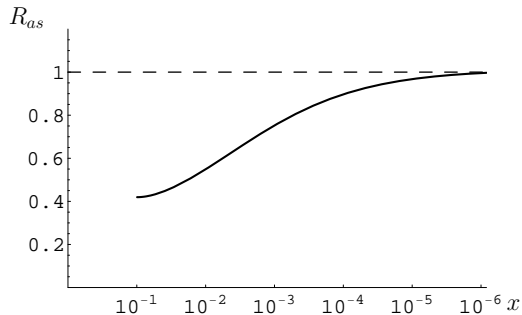}
  \caption{\label{f1dlfig2} Approach of $A_q$ to its asymptotics
$A_q^{as}$ at fixed $Q^2=\mu^2$.}
\end{figure}

Fig.~2 demonstrates that $R_{as} = 0.9$ at $x \approx 8. 10^{-5}$
while the curve in Fig.~3, where $Q^2 = 100 \mu^2$, grows slower and achieves the value $R_{as} = 0.9$ much later, at $x \approx 3.10^{-7}$:

\begin{figure}[h]
  \includegraphics[width=.4\textwidth]{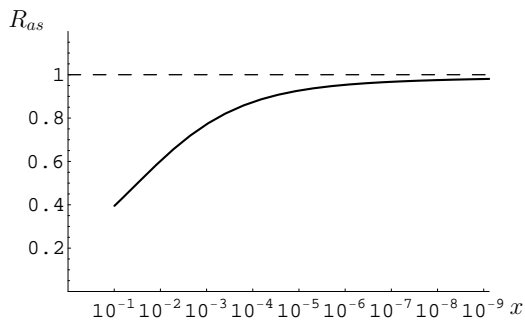}
  \caption{\label{f1dlfig3} Approach of $A_q$ to its asymptotics
$A_q^{as}$ at fixed $Q^2=100\mu^2$~.}
\end{figure}

Therefore, the applicability region of the small-$x$ asymptotics essentially depends on the $Q^2$ value.
The plots in Figs.~2,3 lead us to conclude that the small-$x$ asymptotics reliably represent $F_1$ in the wide
range of $Q^2$ when $x < x_{max}$, with $x_{max} \approx 10^{-6}$.

\section{Summary and outlook}

In this paper we have calculated the perturbative contributions $F_1^q$ and $F_1^g$ to the structure function $F_1$ in the Double-Logarithmic Approximation,
by collecting the DL contributions and at the same time accounting for the running $\alpha_s$ effects.
We obtained the explicit expressions for $F_1^{q,g}$  and then, applying the saddle-point method, calculated the small-$x$ asymptotics of $F_1$,
arriving at the new, DL contribution to the QCD Pomeron. We demonstrated that despite the lack of the factor $1/x$ in the DL contributions,
the impact of their total resummation makes this Pomeron be supercritical, albeit the value of the intercept strongly depends on the
accuracy of calculations.
The maximal value of the intercept
corresponds to the roughest approximation where quark contributions are neglected and $\alpha_s$ is fixed.
Then, the value of the intercept decreases when accuracy of the calculations increases: first,
when the quark contributions are accounted for and then, notably, when the running
$\alpha_s$ effects are taken into account. Nevertheless, the Pomeron remains supercritical as $\omega_0 = 1.066$.
Such monotonic decrease allows us suggest that further accounting
for sub-leading contributions
can decrease the value of the intercept down to zero, so that eventually the intercept will satisfy the Froissart bound.
We proved that
 the $x$ and $Q^2$ -dependencies of $F_1$ converge at small $x$ in dependence on the single variable $Q^2/x^2$.
 We call this convergence the asymptotic scaling. We stress that this prediction of the asymptotic scaling can be
 confirmed by analysis of available experimental data.
 As asymptotically $F_2 \sim 2x F_1$, the
  asymptotic scaling should also take place for $F_2$.
Investigating the applicability region for the asymptotics,
we found that $F_1$  can reliably be represented by its asymptotics at $x \leq x_{max}$, with $x_{max} \approx 10^{-6}$.\\

Although we have discussed the structure function $F_1$, we would like to notice that the  experimental date available in the literature are mostly on the
structure functions $F_2$ and $F_L$, so  it would be interesting to apply our approach to
calculate $F_2$ and $F_L$ as well. Calculating $F_2$ in DLA can be done in the way quite similar to that we have
used for $F_1$.
As a result, we obtain that $F_2$ in DLA can be represented through $F_1$:

\begin{equation}\label{f2dl}
F_2 = 2 x F_1,
\end{equation}
which coincides with the well-known Born relation between $F_1$ and $F_2$.
Eq.~(\ref{f2dl}) entails that in DLA
$F_L = 0$. In order to estimate deviation of
$F_L$ from zero, one should account for sub-leading contributions to both $F_1$ and $F_2$.
In the first place, such contributions are the single-logarithmic (SL) ones.
In this regard we remind that the SL contributions to $F_2$ following from emission of
gluons with momenta widely separated in rapidity and not ordered in transverse momenta
were accounted in Refs.~\cite{catciaf}-\cite{haut}, which involved dealing with the BFKL
characteristic function. However, there are the SL contributions unrelated to BFKL, i.e., for the case of $F_1$,
the SL terms unaccompanied by the factor $1/x$ similarly to the DL terms in Eq.~(\ref{dlser}).
In contrast to the DL contributions (\ref{dlser}), there is not a general technology in the literature for
 resummations of such non-BFKL SL terms.  On the other hand, we were able to modify the IREE method
for the spin structure function $g_1$ (see Ref.~\cite{egtg1sum} and refs therein)
to account for the SL contributions  which are complementary to the  ones calculated in Refs.~\cite{catciaf}-\cite{haut}
namely, the SL following from emission of the partons with momenta ordered in the $k_{\perp}$-space and disordered in
the longitudinal space.
We plan to
adapt this approach to calculate the SL contributions to $F_{1,2}$.
\\

Finally, we stress that in contrast to DGLAP we do not need singular factors $\sim x^{-a}$ in
fits for the initial parton distributions for $F_1$. Such factors  cause a steep rise of the structure functions
at small $x$ and lead to the Regge asymptotics of $F_1$. However, we have shown in Sect.~IV that the resummation of the DL
contributions to $F_1$ automatically leads to the Regge asymptotics, which makes unnecessary inclusion of the singular terms into
the fits.
This result agrees with our earlier results (see Ref.~\cite{egtg1sum}) for the
structure functions $g_1$ and $F_1$ non-singlet and also agrees with the
 results of Refs.~\cite{ehb,haut} obtained for the small-$x$ behavior of the structure function $F_2$. The latter agreement is
 especially interesting because approaches used in Refs.~\cite{ehb,haut} and in the present paper are 
 totally different.

\section{Acknowledgement}

We are grateful to Mario~Greco for interesting discussions of the running $\alpha_s$ effects. Work of B.I.~Ermolaev was
supported in part by the RFBR  Grant No. 16-02-00790-a.

\end{document}